\documentclass[3p,times,twocolumn]{elsarticle}

\usepackage{ecrc_mod}

\volume{-- Preprint}

\firstpage{1}

\journalname{Nuclear Physics B Proceedings Supplement}

\runauth{R.~B.~Patterson}

\jnltitlelogo{Nuclear Physics B Proceedings Supplement}

\usepackage{amssymb}

\usepackage[figuresright]{rotating}

\newcommand\numunue{\ensuremath{\nu_{\mu}\,\mathord{\rightarrow}\,\nu_{e}}}
\newcommand\numunumu{\ensuremath{\nu_{\mu}\,\mathord{\rightarrow}\,\nu_{\mu}}}

\newcommand\anumuanue{\ensuremath{\overline{\nu}_{\mu}\,\mathord{\rightarrow}\,\overline{\nu}_{e}}}
\newcommand\anumuanumu{\ensuremath{\overline{\nu}_{\mu}\,\mathord{\rightarrow}\,\overline{\nu}_{\mu}}}

\newcommand{\numu}{\ensuremath{\nu_{\mu}}}                   
\newcommand{\nue}{\ensuremath{\nu_{e}}}                      
\newcommand{\nutau}{\ensuremath{\nu_{\tau}}}                 
\newcommand{\anu}{\ensuremath{\overline{\nu}}}               
\newcommand{\anue}{\ensuremath{\overline{\nu}_{e}}}          
\newcommand{\anumu}{\ensuremath{\overline{\nu}_{\mu}}}       

\newcommand{\nova}{NOv$_{\!}$A}                                 

\newcommand{\etal}{{\it et al.}}

\begin{document}

\begin{frontmatter}



\dochead{}

\title{The \nova{} experiment:\ status and outlook}

\author[label1]{R.~B.~Patterson\fnref{collab}}
\fntext[collab]{On behalf of the \nova{} collaboration.}
\address[label1]{California Institute of Technology, Pasadena, California 91101}

\begin{abstract}
The \nova{} long-baseline neutrino oscillation experiment is currently under construction and will use an upgraded NuMI neutrino source at Fermilab and a 14-kton detector at Ash River, Minnesota to explore the neutrino sector. \nova{} uses a highly active, finely segmented detector design that offers superb event identification capability, allowing precision measurements of \nue{}\,/\,\anue{} appearance and \numu{}\,/\,\anumu{} disappearance, through which \nova{} will provide constraints on $\theta_{13}$, $\theta_{23}$, $|\Delta m^2_{\mathrm{atm}}|$, the neutrino mass hierarchy, and the CP-violating phase $\delta$.  In this article, we review \nova{}'s uniquely broad physics scope, including sensitivity updates in light of the latest knowledge of $\theta_{13}$, and we discuss the experiment's construction and operation timeline.

\end{abstract}

\end{frontmatter}


\section{Introduction}\label{sec:introduction}

The NuMI Off-axis \nue{} Appearance (\nova{}) experiment is a two-detector, long-baseline, atmospheric-regime neutrino oscillation experiment designed to address a broad range of open questions in the neutrino sector through precision measurements of \numunue{}, \anumuanue{}, \numunumu{}, and \anumuanumu{} oscillations.  As the name implies, much of \nova{}'s physics scope comes from the appearance channels, as the observed rates of \nue{} and \anue{} interactions provide information on (1) the ordering of the neutrino masses ({\it i.e.}, whether the $\nu_3$ state is more or less massive than the other two), (2) the amount of CP violation present in the neutrino sector, (3) the size of the PMNS mixing angle $\theta_{13}$, and (4) whether the $\nu_3$ state has more \numu{} or \nutau{} admixture (that is, whether $\theta_{23}\mathord{>}\frac{\pi}{4}$ or $\mathord{<}\frac{\pi}{4}$, respectively).  Recent results by short-baseline \anue{} disappearance experiments demonstrating $\theta_{13}\,\mathord{\approx}\,9^{\circ}$~\cite{doublechooz,dayabay,reno} ensures that \nova{} will have substantial event rates in the appearance channels.  Through \numu{} and \anumu{} disappearance, \nova{} will provide improved precision on the dominant atmospheric oscillation parameters $\theta_{23}$ and $\left|\Delta m_\mathrm{atm}^2\right|$.

Outside of these primary goals, \nova{} will also look for evidence of new physics through comparisons of \numunumu{} and \anumuanumu{}, provide constraints on sterile neutrino models by measuring the total flux of active neutrinos at its downstream detector, monitor for supernova neutrino activity, perform neutrino-nucleus cross section measurements with a narrow-band beam, and pursue a variety of non-neutrino topics including searches for magnetic monopoles and hidden sector particles.

For all of the oscillation measurements, \nova{} takes advantage of a two-detector configuration to mitigate uncertainties in neutrino flux, neutrino cross sections, and event selection efficiencies.  The 14-kton Far Detector (FD) is currently under construction in Ash River, Minnesota, 810 km downstream of the neutrino source at Fermilab.  The 0.3-kton Near Detector (ND) will be located on the Fermilab site in a new cavern to be excavated near the existing MINOS Near Detector Hall.

\section{Neutrino source}
\begin{figure}[bt]
\begin{center}
	\resizebox{\linewidth}{!}{\includegraphics[viewport=0 0 522 410, clip=true]{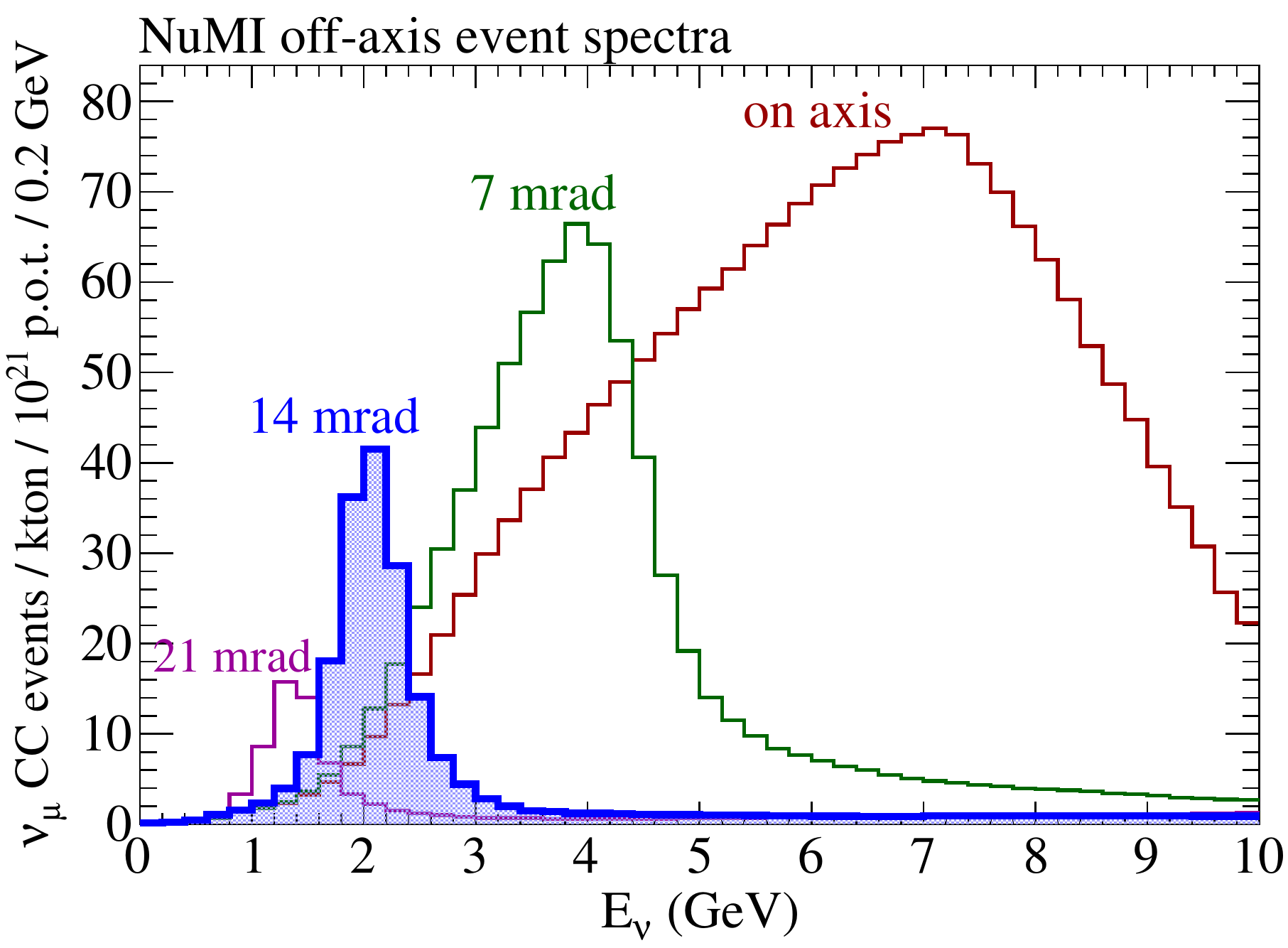}}
	\caption{Simulated neutrino energy spectra for \numu{} charged current interactions in detectors sited 0, 7, 14, and 21 mrad off the NuMI beam axis.  \nova{} sits at 14 mrad.}
	\label{fig:spectrum}
\end{center}
\end{figure}

\nova{} uses Fermilab's NuMI beamline as its neutrino source~\cite{numi}.  The \nova{} detectors are situated 14 mrad off the NuMI beam axis, so they are exposed to a relatively narrow band of neutrino energies centered at 2~GeV.  Figure~\ref{fig:spectrum} shows how the energy spectrum for \numu{} charged current (CC) events varies with detector position.  The suppressed high-energy tail at \nova{}'s off-axis location reduces neutral current backgrounds in the visible energy range of 1 to 3~GeV where the appearance of \nue{} CC events should occur.

The NuMI source is undergoing upgrades to increase its average beam power from 350~kW to 700~kW.  Much of the increased power comes from a reduction in the Main Injector cycle time, which will drop from 2.2 seconds to 1.3 seconds.  This cycle time reduction is in turn made possible by reconfiguring the antiproton Recycler as a proton injection ring, thereby allowing ramping in the Main Injector to occur concurrently with the next injection.  The NuMI upgrades are scheduled to last 12 months, ending May 2013.

\section{Detectors}
\begin{figure}
\begin{center}
	\resizebox{\linewidth}{!}{\includegraphics[viewport=10 0 550 370, clip=true]{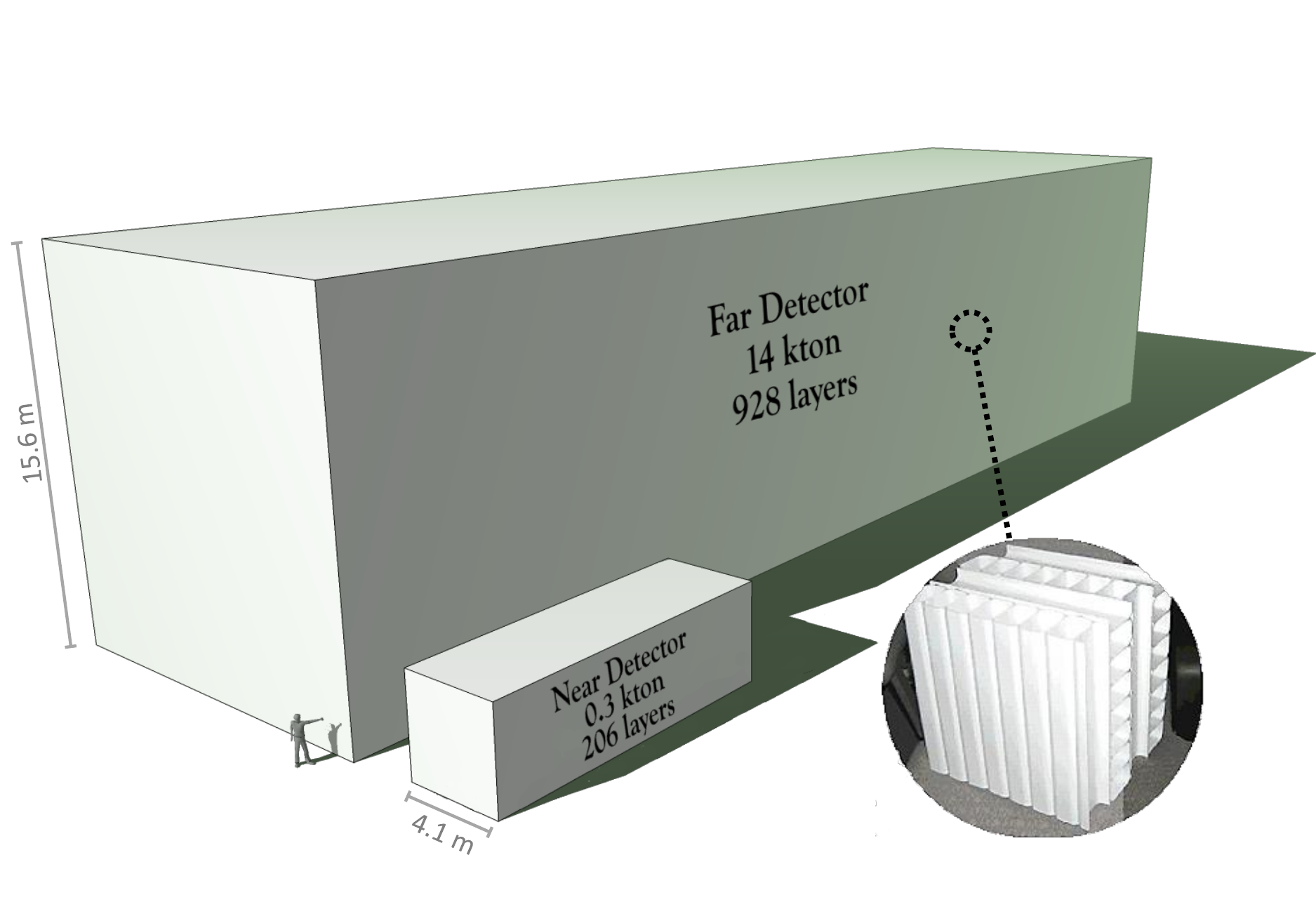}}
	\caption{\nova{} detectors, with a human figure shown for scale.  The FD differs from the ND only in the length of its PVC cells and the number of layers present.  Each layer in the detectors is oriented orthogonally to adjacent ones to provide 3D event reconstruction.  ({\it Inset}) A cut-away view of the PVC cellular structure.}
	\label{fig:detectors}
\end{center}
\end{figure}

The \nova{} detectors are highly segmented, highly active tracking calorimeters.  The segmentation and the overall mechanical structure of the detectors are provided by a lattice of PVC cells with cross sectional size (6~cm)$\times$(4~cm).  Each cell extends the full width or height of the detector -- 15.6~m in the FD, 4.1~m in the ND -- and is filled with liquid scintillator.  Light produced by the scintillator is collected and transported to the end of the cell by a wavelength-shifting fiber that terminates on a pixel of a 32-channel avalanche photodiode.  Figure~\ref{fig:detectors} shows a sketch of the FD and ND along with a cut-away view of the PVC lattice.  Each of the 928 layers of the FD has 384 cells, for $\sim$360,000 total channels of readout.  The ND has 206 layers each with 96 cells plus a muon range stack at the downstream end (not shown in the figure) made by interleaving steel plates with standard detector layers.  

\begin{figure}[hbt]
\begin{center}
	\resizebox{\linewidth}{!}{\includegraphics[viewport=0 0 548 456, clip=true]{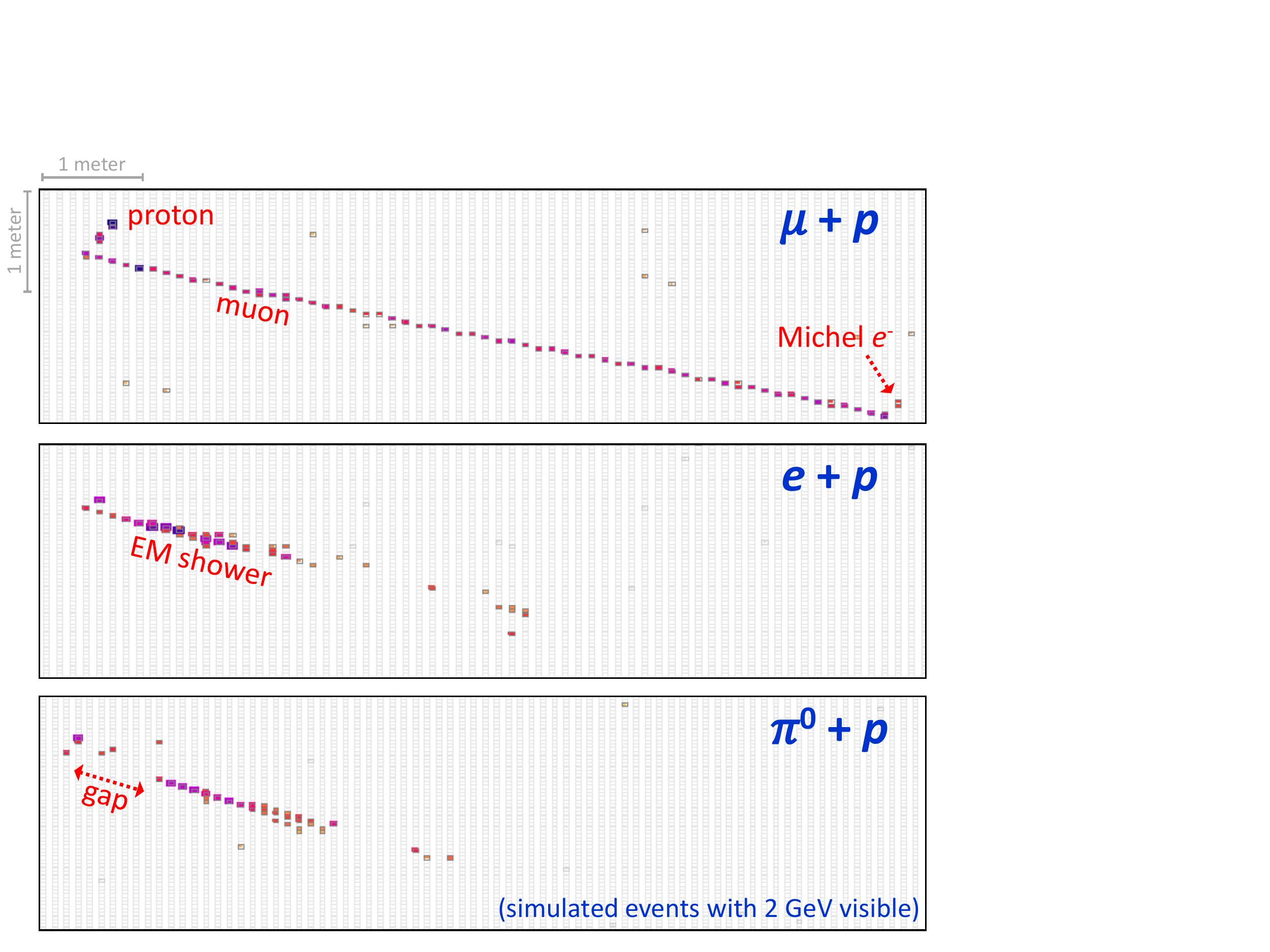}}
	\caption{Simulated ND events with 2~GeV visible energy.  Each colored rectangle corresponds to a single detector cell, with darker colors representing higher measured energy deposition. Only one of the two orthogonal detector views is shown for each event.  ({\it Top}) \numu{} CC quasi-elastic event, with a muon and proton in the final state.  The muon's decay electron is also visible at the end of the track.  While it is included in the display here, this delayed activity can be temporally separated at analysis time.  ({\it Middle}) \nue{} CC quasi-elastic event.  ({\it Bottom}) NC resonant $\pi^0$ production, with a $\pi^0$ and proton in the final state.  One decay photon carries the majority of the $\pi^0$'s momentum, and its activity is clearly separated from the interaction point.}
	\label{fig:events}
\end{center}
\end{figure}

Figure~\ref{fig:events} shows three simulated events in the \nova{} ND.  Muons are clearly identifiable as long, straight tracks with appropriate energy deposition per unit pathlength ($\frac{dE}{dx}$).  Proton tracks can be separated from other hadron tracks by their $\frac{dE}{dx}$ profiles.  The \nova{} detector technology is particularly well-suited for electromagnetic shower identification, as the radiation length in the detector (38~cm) is many times larger than the relevant PVC cell dimensions.  This level of granularity helps $\pi^0$ decays stand out, as the decay photons leave telltale gaps in detector activity between the neutrino interaction location and the photon conversion point, as in the bottom panel of Figure~\ref{fig:events}.

\begin{figure}[hbt]
\begin{center}
	\resizebox{\linewidth}{!}{\includegraphics[viewport=0 0 281 505, clip=true]{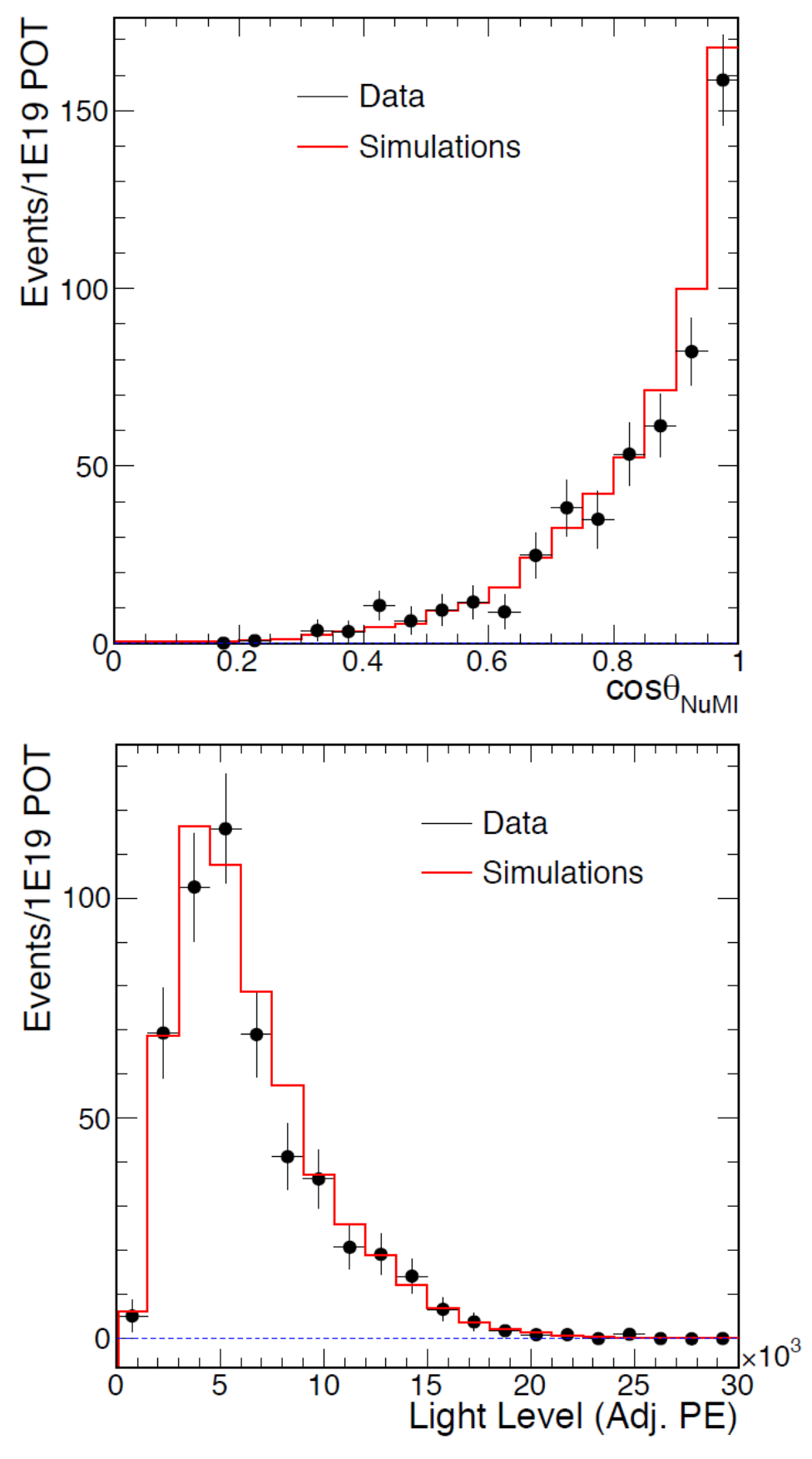}}
	\caption{NuMI neutrino events in the NDOS, data and simulation.  ({\it Top}) Reconstructed direction of the longest track relative to the nominal beam direction. ({\it Bottom}) Total observed energy in the event, shown here in units of ``adjusted photoelectrons''.   Cosmic ray backgrounds, measured using beam-off data, have been subtracted from both distributions.}
	\label{fig:ndosdatamc}
\end{center}
\end{figure}

Since November 2010, \nova{} has operated a prototype detector, dubbed the Near Detector on the Surface (NDOS), that has allowed full-scale detector assembly and integration tests, electronics and data acquisition development, calibration R\&D, Monte Carlo simulation tuning, and early analysis R\&D.  The NDOS sits 110~mrad off the NuMI beam axis and approximately on the Booster beam axis and is identical in size to the ND except in its width, with 64 cells spanning it horizontally rather than 96.  With the NDOS, \nova{} has recorded hundreds of neutrino interactions from both the NuMI and Booster sources and has collected millions of cosmic ray interactions.  Figure~\ref{fig:ndosdatamc} shows two distributions using NuMI neutrino events:\ the reconstructed primary track direction and the total visible energy.  The data and simulation are in excellent agreement.

\begin{figure}[hbt]
\begin{center}
	\resizebox{\linewidth}{!}{\includegraphics[]{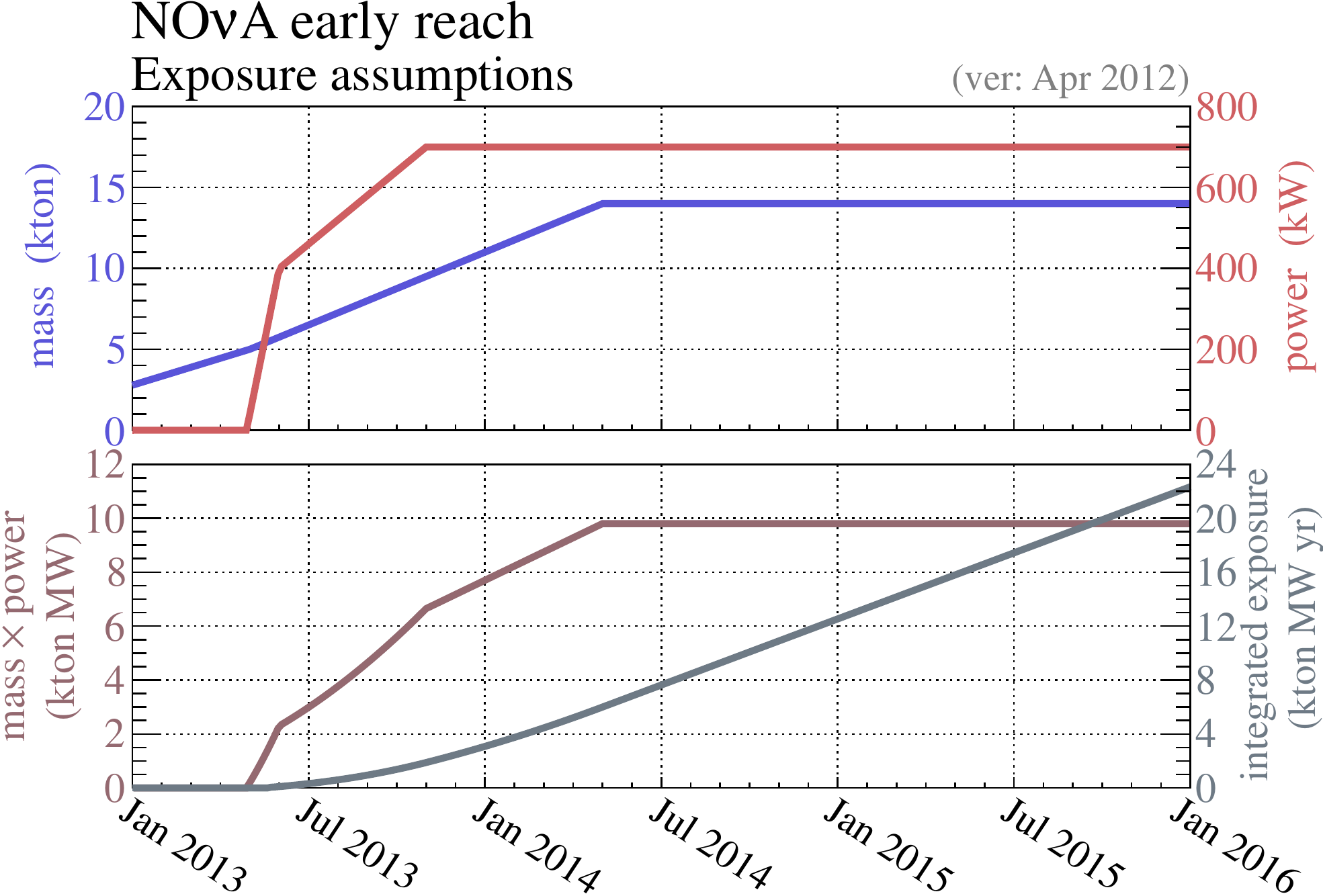}}
	\caption{Expected development of detector mass, beam power, instantaneous exposure, and integrated exposure in the first few years of \nova{} running.}
	\label{fig:exposure}
\end{center}
\end{figure}

The NDOS continues to operate.  FD construction has begun and will proceed over a two-year period ending Summer 2014.  Neutrino data taking will commence with a partial detector ($\sim$5~kton) when the NuMI beam returns in May 2013.  After a six-month beam commissioning period, the NuMI line will be ready for its full 700-kW operation.  Figure~\ref{fig:exposure} shows the expected \nova{} exposure during early running.  

\section{\nue{}\,/\,\anue{} appearance}
\begin{figure}
\begin{center}
	\resizebox{\linewidth}{!}{\includegraphics[]{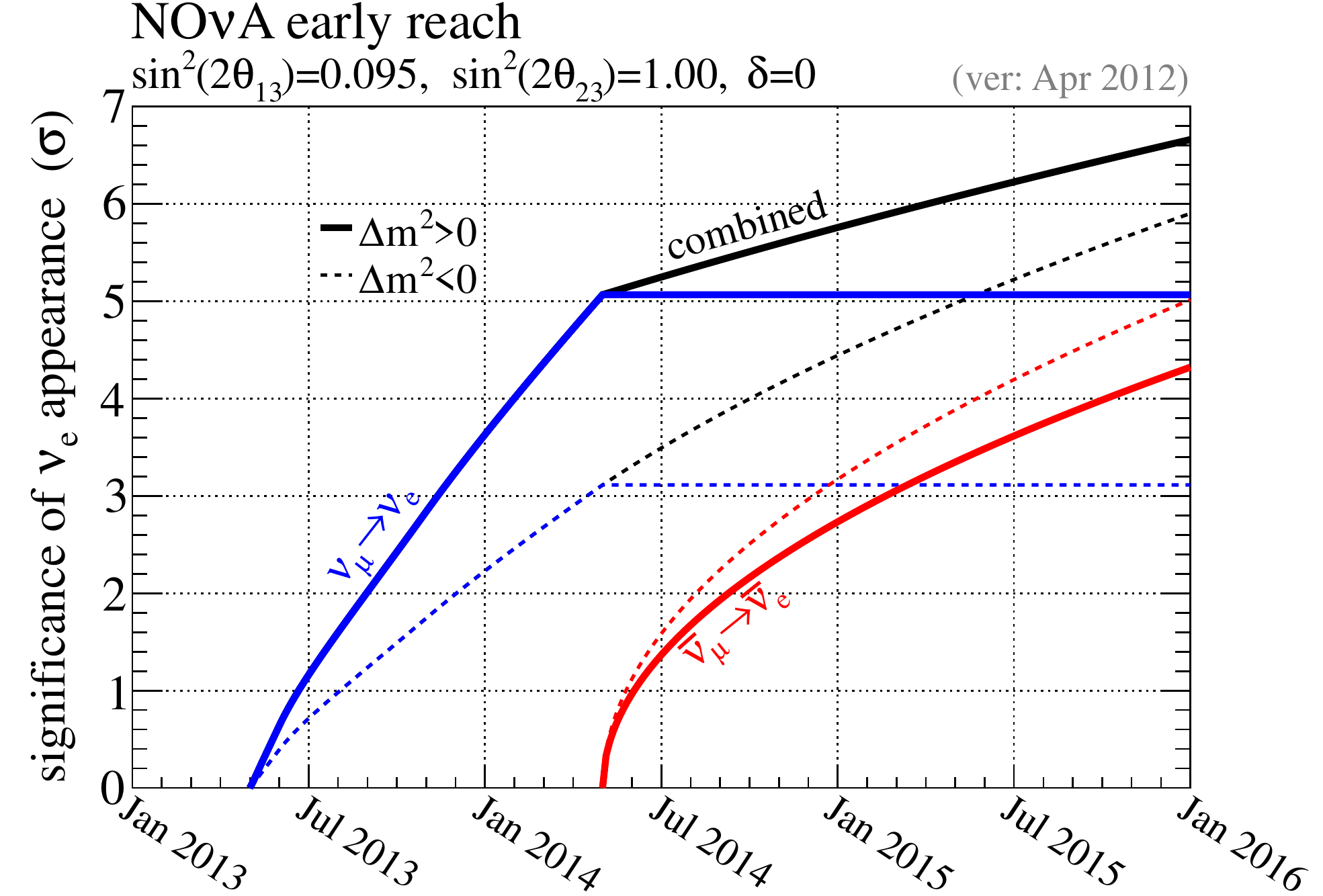}}
	\caption{Sensitivity to \numunue{} and \anumuanue{} oscillations in early running.  The solid (dashed) lines assume the normal (inverted) hierarchy.  The red curves that begin May 2014 assume a switch to \anu{} running after one year.  This run plan is merely an example, and the time of switching to \anu{} running is flexible.  The black curves labeled ``combined'' show the appearance significance for the combined \numunue{} and \anumuanue{} samples.}
	\label{fig:earlyreach}
\end{center}
\end{figure}

\nova{} will begin its oscillation data run in $\nu$ mode, that is with the NuMI horn configured to focus positive secondary hadrons.  While FD construction and NuMI commissioning will still be underway during the first year of data taking, \nova{} can nonetheless reach a 5$\sigma$ C.L.\ observation of $\theta_{13}$-driven \numunue{} oscillations after one year, assuming the normal mass hierarchy.  Figure~\ref{fig:earlyreach} shows how this sensitivity changes with time.

The baseline \nova{} exposure is $3.6\mathord{\times}10^{21}$~protons-on-target (p.o.t.),\  which can be accumulated in six years at design intensities, with a 14-kton FD.  For the sensitivities shown below, this exposure is assumed divided evenly between $\nu$ and \anu{} running, a split that works well for many parameter scenarios but is nonetheless adjustable.  The analysis techniques used here are those described in the \nova{} Technical Design Report~\cite{tdr}.  Updated analyses for use in the first \nova{} results are under active development.

Table~\ref{tab:events} shows the number of \nue{} and \anue{} CC event candidates expected after a six-year run.  Since these counts depend strongly on the exact oscillation parameters, a sort of ``average'' is shown here by assuming no matter effects, no solar oscillations, and $\theta_{23}\mathord{=}\frac{\pi}{4}$.  The signal efficiency is $\sim$45\% and the neutral current (NC) leakage rate is $\sim$1\%.  These rates are somewhat higher than those in the Technical Design Report as the analysis cuts have been reoptimized for a non-zero value of $\theta_{13}$~\cite{tdr}.

\begin{table}
\begin{center}
\begin{tabular}{r|cc}
&$\nu$&$\overline{\nu}$\\
\hline
NC&19&10\\
\numu{} CC&5&$\mathord{<}1$\\
\nue{} CC&8&5\\
\hline
total bg.&32&15\\
\hline
\numunue{} CC&68&32
\end{tabular}
\caption{Representative numbers of \numunue{} and \anumuanue{} candidate events expected in the FD after three years each of $\nu$ and \anu{} running, assuming $\sin^2(2\theta_{13})\mathord{=}0.095$.  The two columns are labeled by running mode, $\nu$ or \anu{}.  Each row includes both right- and wrong-sign components.}
\label{tab:events}
\end{center}
\end{table}

\begin{figure}[hbt]
\begin{center}
	\resizebox{\linewidth}{!}{\includegraphics[viewport=10 5 528 494, clip=true]{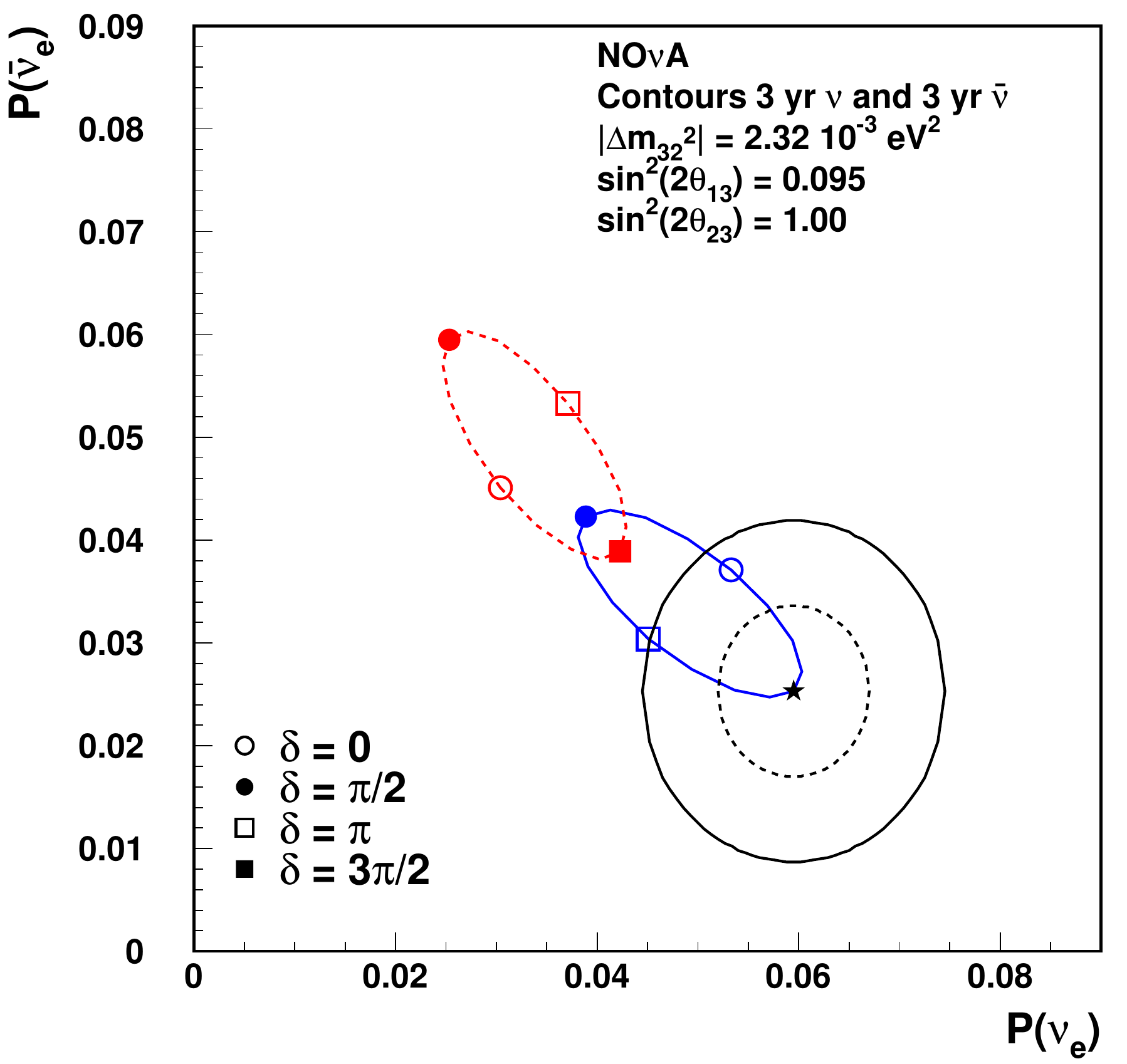}}
	\caption{An overview of the \nova{} appearance measurements.  For $\sin^2(2\theta_{13})\,\mathord{=}\,0.095$ and $\sin^2(2\theta_{23})\,\mathord{=}\,1$, all possible values for the \mbox{2-GeV} appearance probabilities $P(\nue{})$ and $P(\anue{})$ are shown.  The solid blue (dashed red) ellipse corresponds to the normal (inverted) hierarchy scenarios, with $\delta$ varying as one moves around each ellipse.  1$\sigma$ and 2$\sigma$ sensitivities to $P(\nue{})$ and $P(\anue{})$ are shown in black for the test case at  $\delta\,\mathord{=}\,\frac{3\pi}{2}$, normal hierarchy (starred point).}
	\label{fig:basicbiprob}
\end{center}
\end{figure}

\begin{figure}
\begin{center}
	\resizebox{\linewidth}{!}{\includegraphics[viewport=5 0 520 390, clip=true]{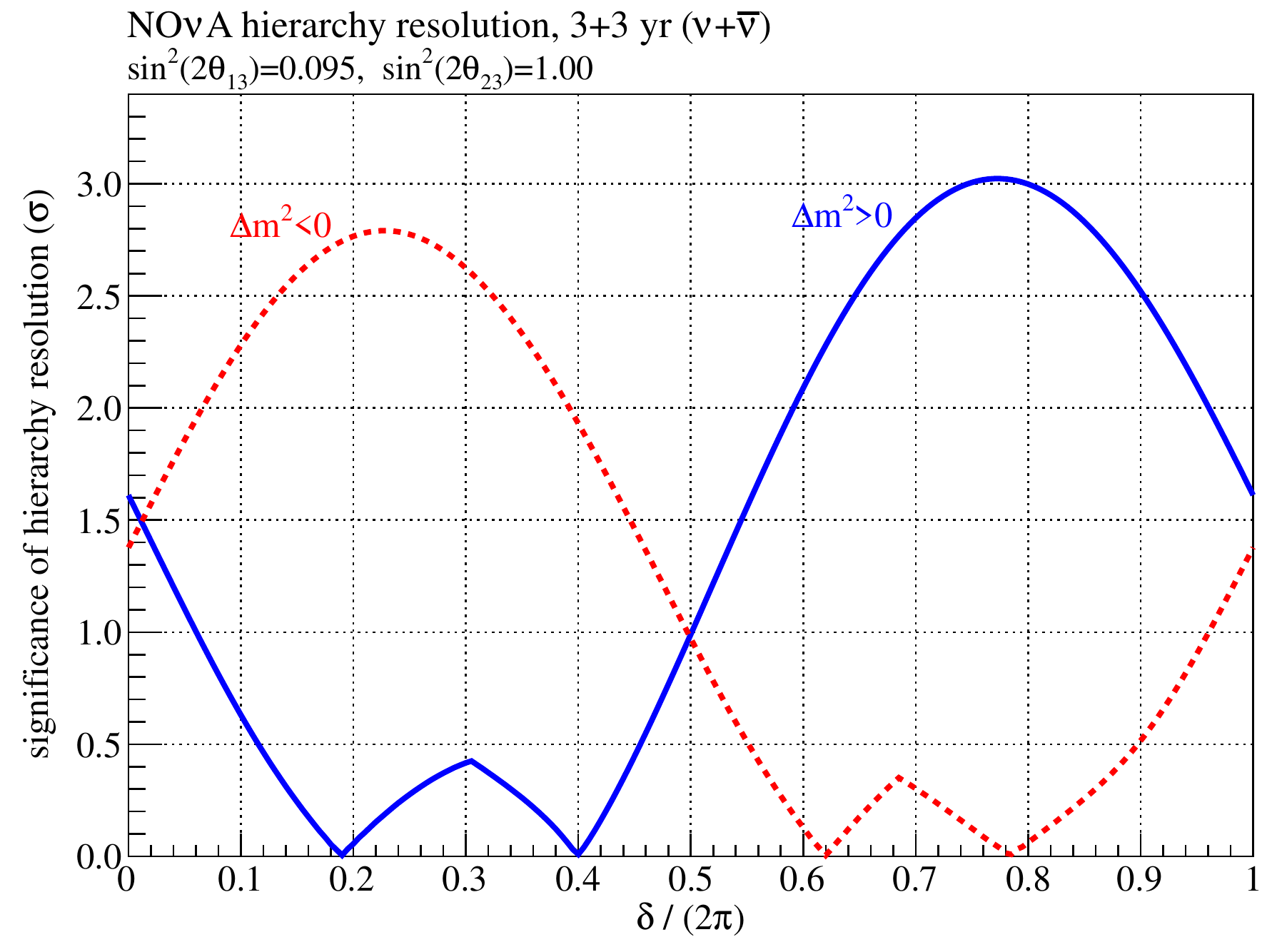}}
	\caption{Significance with which \nova{} can resolve the mass hierarchy as a function of $\delta$ for the indicated values of $\sin^2(2\theta_{13})$ and $\sin^2(2\theta_{23})$.  $3.6\mathord{\times}10^{21}$~p.o.t.,\ evenly split between $\nu{}$ and $\anu{}$ running, is assumed. The solid blue (dashed red) curve shows the expected significance assuming the normal (inverted) hierarchy.}
	\label{fig:hier}
\end{center}
\end{figure}

Given its narrow-band beam, \nova{} primarily measures oscillations of 2~GeV neutrinos.  Thus, the appearance measurement approximately reduces to the extraction of two numbers from the data:\ the probabilities for \numunue{} and \anumuanue{} oscillations at 2~GeV.  These two probabilities depend on both the mass hierarchy and the value of the CP-violating phase $\delta$, and it is instructive to see these dependencies graphically.  Figure~\ref{fig:basicbiprob} shows the locus of possible values for these two probabilities for a given set of mixing angles.  Overlayed on the figure are the expected 1$\sigma$ and 2$\sigma$ C.L.\ contours for a particular test point after six years of \nova{} running.  This representation of the appearance measurements, though approximate, makes plain the nature of \nova{}'s hierarchy and CP violation sensitivity.  For example, no possible inverted hierarchy scenarios are included in the 2$\sigma$ intervals for the test point shown, so the inverted hierarchy would be excluded by at least 2$\sigma$ in this scenario.\footnote{A statistical note:\ the confidence levels shown in this figure assume one free parameter, with 2$\sigma$ corresponding to $-2\Delta\log \cal{L}\,\mathord{=}\,\mathrm{4}$.  This is appropriate given that the 2D space $\left\{P(\nue{}),P(\anue{})\right\}$ is not actually dense with possible answers.  One can also read the contours as representing the $-2\Delta\log\cal{L}$ surface itself, allowing comparisons between the test point and any alternative allowed hypothesis.  In either case, this figure is intended to provide an intuitive picture of the measurement principle, the sensitivity of which is quantified more transparently elsewhere.}  Figure~\ref{fig:basicbiprob} also demonstrates that the significance with which \nova{} can establish the hierarchy depends on $\delta$.  This significance is shown explicitly as a function of $\delta$ in Figure~\ref{fig:hier}.

\begin{figure}[hbt]
\begin{center}
	\resizebox{\linewidth}{!}{\includegraphics[viewport=5 0 520 390, clip=true]{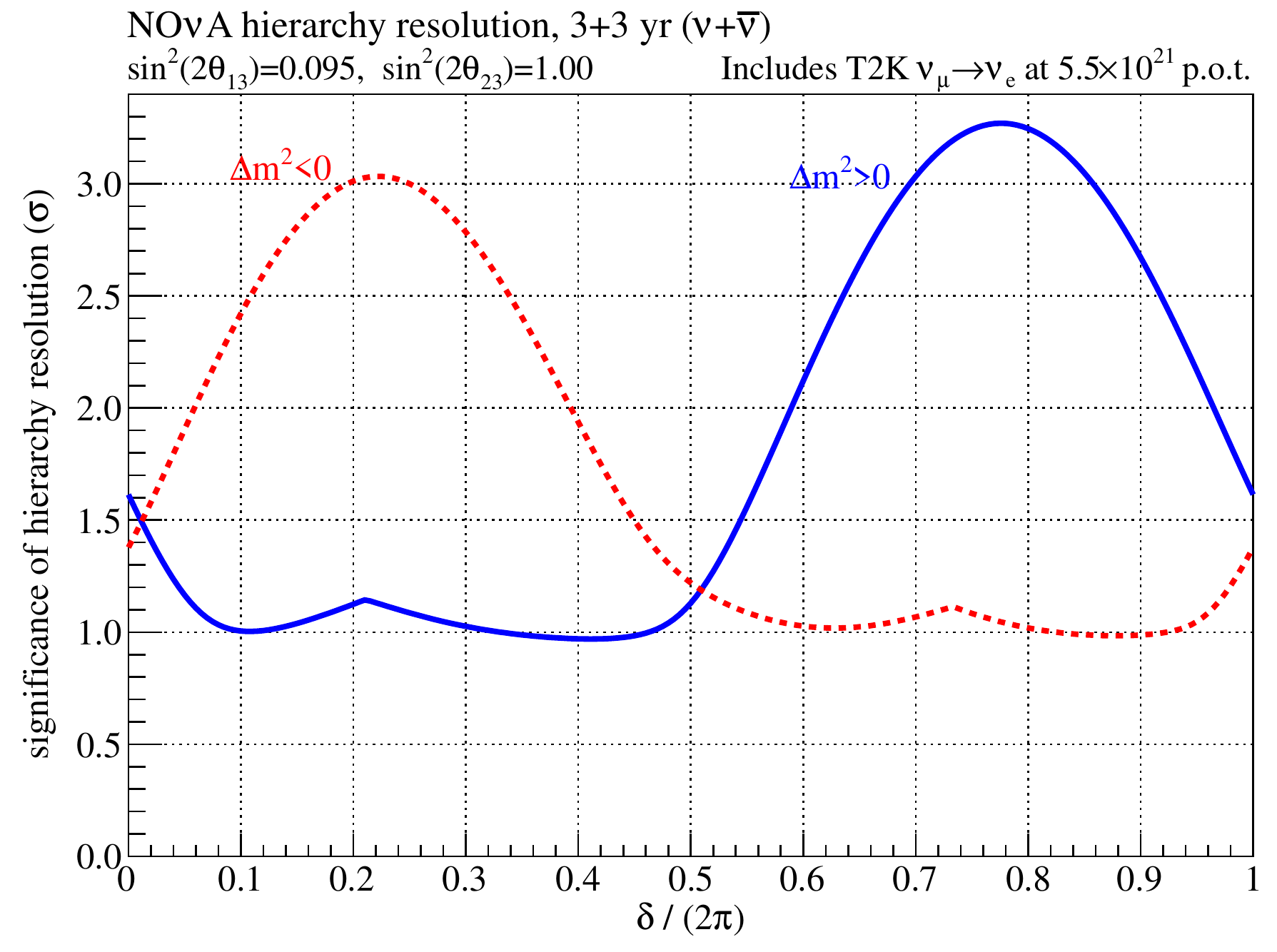}}
	\caption{A version of Figure~\ref{fig:hier} that includes $5.5\mathord{\times}10^{21}$~p.o.t.\ of T2K \numunue{} data.  The T2K data helps lift the degeneracy in unfavorable hierarchy/$\delta$ scenarios.}
	\label{fig:hier_t2k}
\end{center}
\end{figure}

By the end of the primary \nova{} run, the T2K experiment will have a significant \numunue{} oscillation data set of its own, and the appearance probabilities for T2K depend relatively little on the mass hierarchy~\cite{t2k}.  Thus, potential degeneracies in the \nova{} measurement can be partially lifted by including T2K data.  The combined sensitivity is shown in Figure~\ref{fig:hier_t2k}.

\begin{figure}
\begin{center}
	\resizebox{\linewidth}{!}{\includegraphics[viewport=10 5 528 494, clip=true]{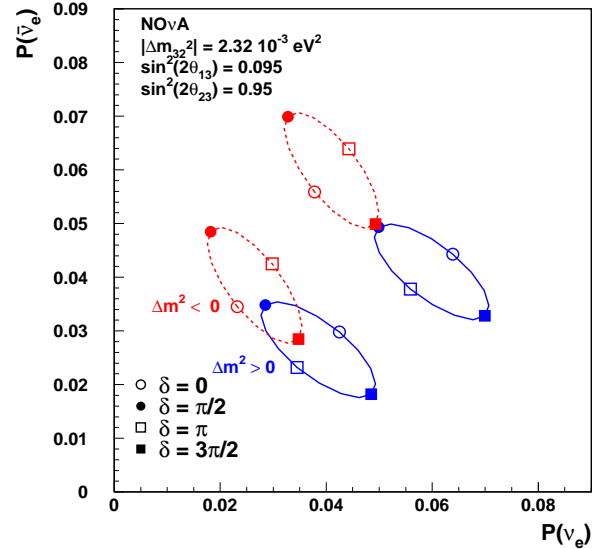}}
	\caption{Similar to Figure~\ref{fig:basicbiprob} but with $\sin^2(2\theta_{23})\,\mathord{=}\,0.95$.  The possible values for the appearance probabilities $P(\nue{})$ and $P(\anue{})$ at 2~GeV have split, with the ellipses on the upper right (lower left) corresponding to $\theta_{23}\mathord{>}\frac{\pi}{4}$ ($\theta_{23}\mathord{<}\frac{\pi}{4}$).}
	\label{fig:splitbiprob}
\end{center}
\end{figure}

The examples given so far have assumed \mbox{$\sin^2(2\theta_{23})\,\mathord{=}\,1$}.  \numu{} disappearance measurements in the coming years, including those from \nova{} (see below), will provide increased precision on $\sin^2(2\theta_{23})$, and it is possible that non-maximal mixing will be established.  Figure~\ref{fig:splitbiprob} shows how non-maximal mixing influences the \nova{} appearance measurements.  In particular, the set of $\left\{P(\nue{}),P(\anue{})\right\}$ values that \nova{} can measure at 2~GeV is now described by four ellipses rather than two, with the higher (lower) probability cases corresponding to \mbox{$\theta_{23}\mathord{>}\frac{\pi}{4}$} \mbox{($\theta_{23}\mathord{<}\frac{\pi}{4}$)}.  Equivalently, the higher probability cases are those where the $\nu_3$ state has more \numu{} than \nutau{} admixture, and vice versa.

\begin{figure}[hbt]
\begin{center}
	\resizebox{\linewidth}{!}{\includegraphics[viewport=5 0 520 390, clip=true]{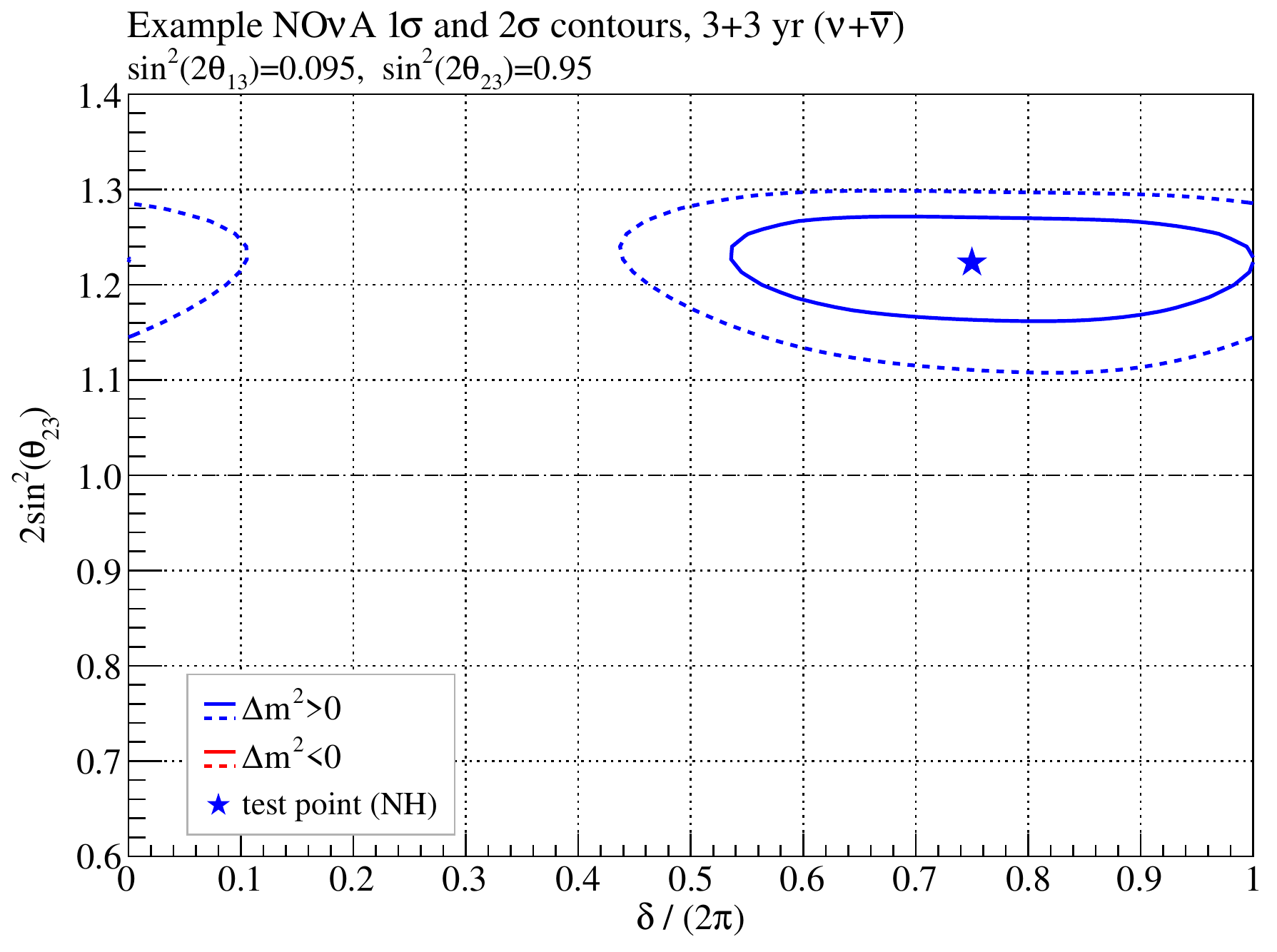}}\vspace{0.25cm}
	\resizebox{\linewidth}{!}{\includegraphics[viewport=5 0 520 390, clip=true]{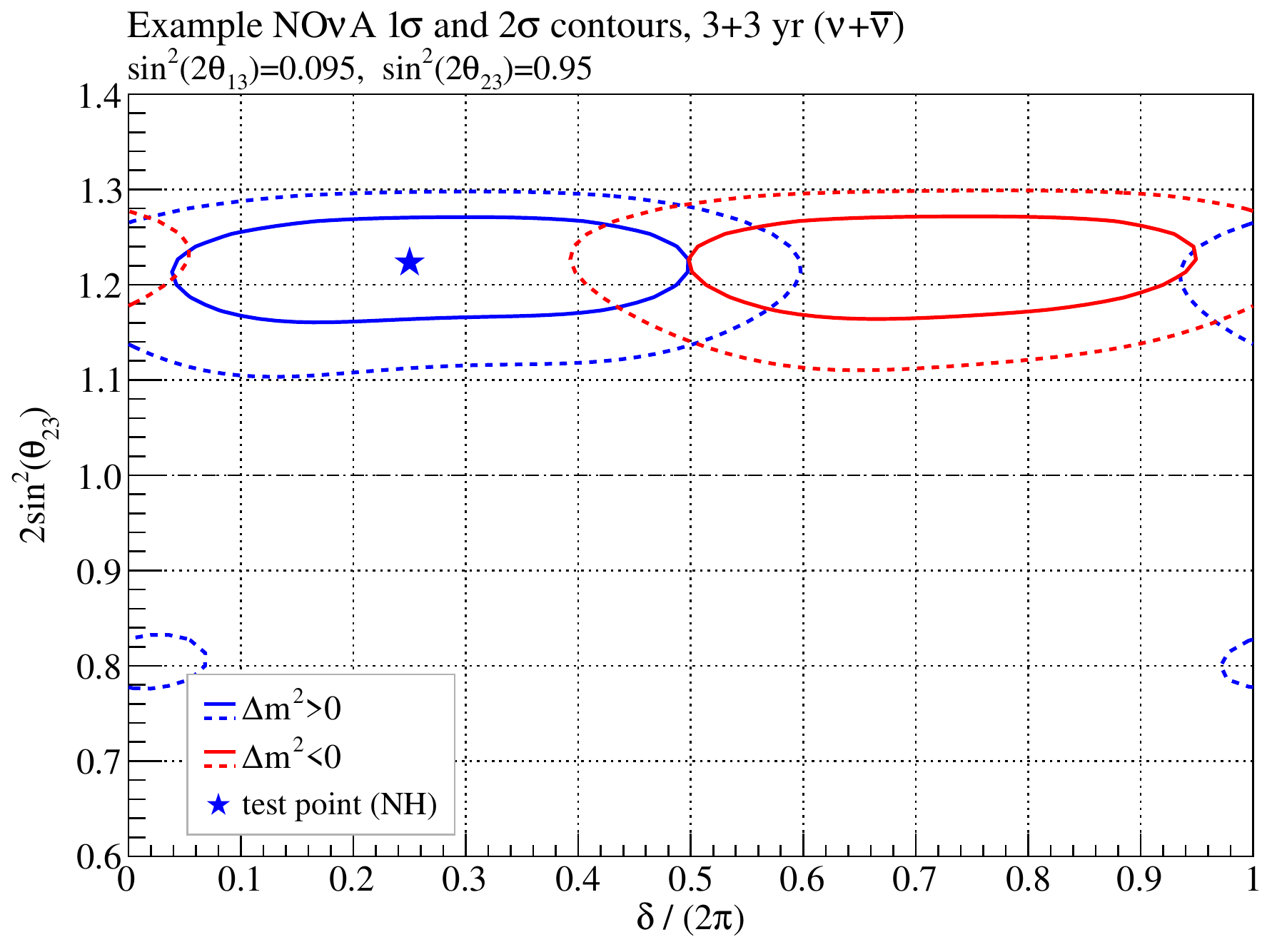}}
	\caption{Example 1$\sigma$ and 2$\sigma$ allowed regions for $\delta$ and $2\sin^2(\theta_{23})$ for the two starred test points.  ({\it Top}) Test point at $\sin^2(2\theta_{13})\,\mathord{=}\,0.095$, $\sin^2(2\theta_{23})\,\mathord{=}\,0.95$, $\theta_{23}\,\mathord{>}\,\frac{\pi}{4}$, $\delta\,\mathord{=}\,\frac{3\pi}{2}$, normal hierarchy (NH). ({\it Bottom}) Same except for $\delta\,\mathord{=}\,\frac{\pi}{2}$.  The blue (red) contours show the expected allowed ranges under a normal (inverted) hierarchy assumption, and the confidence levels are established assuming two free parameters.}
	\label{fig:2dcontour}
\end{center}
\end{figure}

This bifurcation in the set of possible outcomes allows \nova{} to make a measurement of this flavor balance ({\it i.e.}, of the $\theta_{23}$ octant).  If $\sin^2(2\theta_{23})\,\mathord{<}\,1$, then, the \nova{} appearance data will provide information on the mass hierarchy, $\delta$, and the $\theta_{23}$ octant simultaneously.  Figure~\ref{fig:2dcontour} shows \nova{}'s sensitivity to these three unknowns under two example scenarios, one in the degenerate regime and one not.  Several features in the figure deserve mention. (1) The absence of any inverted hierarchy contours in the top panel indicates the $\mathord{>}2\sigma$ significance of hierarchy determination expected in this scenario. (2) The bottom panel shows a ``degenerate'' scenario, in which CP violation and long-baseline matter effects introduce cancelling perturbations in the oscillation probabilities.  In such scenarios, \nova{} provides only correlated information on the mass hierarchy and $\delta$. (3) \nova{}'s $\theta_{23}$ octant sensitivity is largely independent of the hierarchy and $\delta$.  This is seen qualitatively in Figure~\ref{fig:splitbiprob} by the separation of the two octants' probabilities, and it is visible in the bottom panel here, where the correct octant is strongly preferred despite the presence of ambiguities in determining the mass hierarchy and $\delta$.

\subsection{Disappearance}
While electron identification capability was key in the design of \nova{}, the detectors also have excellent energy resolution for \numu{} charged current events, particularly for quasi-elastic interactions, and \nova{} will make precision measurements of the atmospheric oscillation parameters $\sin^2(2\theta_{23})$ and $|\Delta m^2_{\mathrm{atm}}|$ through \numu{} and \anumu{} disappearance.  Like the appearance analyses, the disappearance analyses are under active development, and the sensitivities shown here come from earlier estimates~\cite{tdr} updated to include the latest knowledge of beam and detector performance.

\begin{figure}[hbt]
\begin{center}
	\resizebox{\linewidth}{!}{\includegraphics[viewport=35 30 463 423, clip=true]{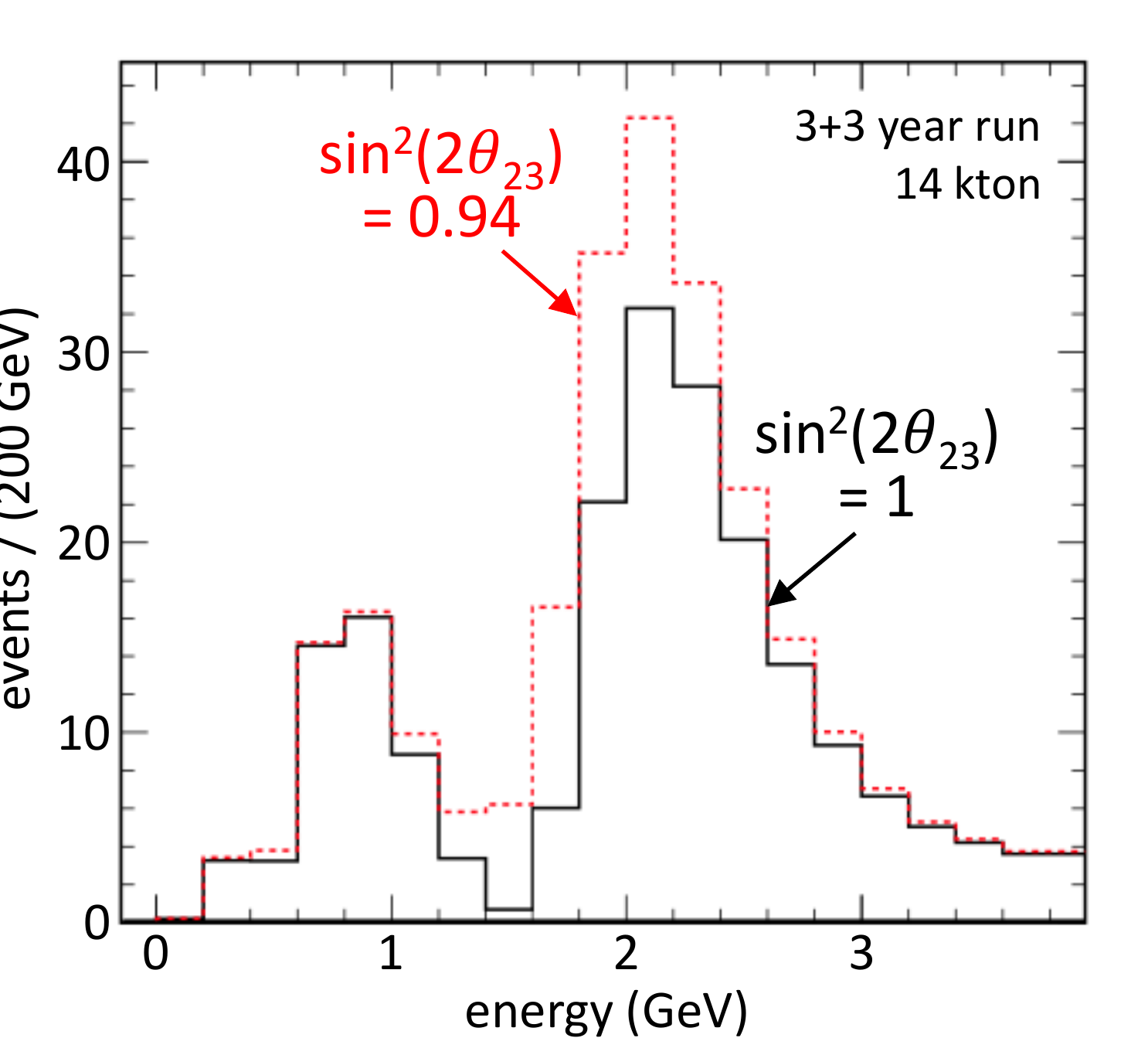}}
	\caption{Reconstructed neutrino energy spectrum expected at the FD in two scenarios after a 3$+$3~year ($\nu\mathord{+}\anu{}$) run.  The solid black (dashed red) histogram is for $\sin^2(2\theta_{23})\,\mathord{=}\,1$ ($\sin^2(2\theta_{23})\,\mathord{=}\,0.94$).}
	\label{fig:numuspectra}
\end{center}
\end{figure}

\nova{}'s narrow-band 2-GeV spectrum and matching 810-km baseline mean that the \numu{} flux is largely oscillated away at the FD.  Figure~\ref{fig:numuspectra} shows the expected reconstructed energy spectrum for \numu{} and \anumu{} CC quasi-elastic events after three years each of $\nu{}$ and \anu{} running at design exposures.  Both a maximal and a non-maximal mixing scenario are shown, and these are readily distinguished given the expected event counts and the energy resolution of 4\% assumed for this quasi-elastic sample.

\begin{figure}[hbt]
\begin{center}
	\resizebox{\linewidth}{!}{\includegraphics[viewport=10 0 565 424, clip=true]{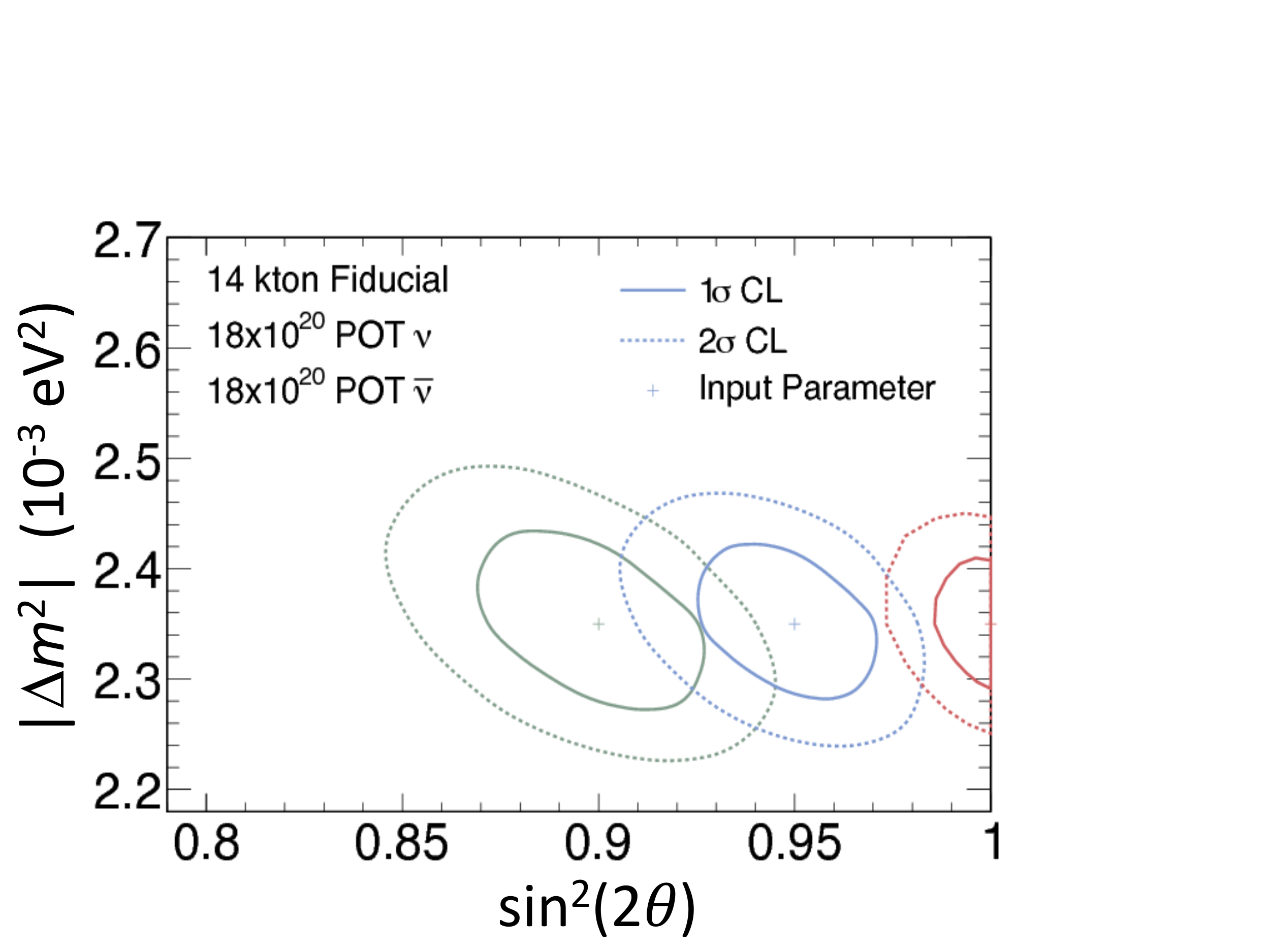}}
	\caption{\nova{} 1$\sigma$ and 2$\sigma$ sensitivity to $\sin^2(2\theta_{23})$ and $|\Delta m^2_{\mathrm{atm}}|$ after a 3$+$3 year ($\nu\mathord{+}\anu{}$) run.}
	\label{fig:numucontour}
\end{center}
\end{figure}

Figure~\ref{fig:numucontour} shows the sensitivity to $\sin^2(2\theta_{23})$ and $|\Delta m^2_{\mathrm{atm}}|$ for three test cases.  If $\sin^2(2\theta_{23})$ is near the current global best-fit value of 0.95~\cite{global}, \nova{} can establish non-maximal mixing at $\mathord{>}2\sigma$.

\section{Closing}

\nova{} FD construction is underway at Ash River, and ND cavern excavation is now beginning at Fermilab.  The ND-sized prototype detector, NDOS, has operated for two years at Fermilab and has allowed rapid development of assembly procedures and analysis tools.

The NuMI beam will return May 2013, and \nova{} will begin collecting neutrino data immediately with a partial detector.  Full instantaneous exposure will be reached in summer 2014 when the 14-kton FD is complete.  With a six-year run, \nova{} can unambiguously resolve the neutrino mass hierarchy at $\mathord{>}95\%$~C.L.\ for over a third of possible values of $\delta$.  \nova{} will otherwise provide $\delta$-dependent hierarchy determination plus measurements of $\theta_{13}$, $\theta_{23}$, $|\Delta m^2_{\mathrm{atm}}|$, and $\delta$ itself.

\nova{} will begin its FD neutrino run in $\nu$ mode, with a future switch to $\anu{}$ running yet to be scheduled.  For the normal mass hierarchy, \nova{} can achieve 5$\sigma$ C.L.\ observation of atmospheric-regime \numunue{} oscillations with its first year of data, even with the exposure limitations of on-going detector construction and NuMI beam commissioning.

\section*{Acknowledgments}
The author acknowledges support from the Department of Energy under contracts ER40701 and ER41735.

\end{document}